\pdfoutput=1

\documentclass[twocolumn, twocolappendix]{aastex63}

\usepackage{amsmath}
\usepackage{comment}
\usepackage{microtype}
\usepackage{natbib}
\setcitestyle{round}
\received{2021 May 5}
\accepted{2021 October 8}
\published{2021 October 19}
\submitjournal{Publications of the Astronomical Society of the Pacific}

\shorttitle{Orbit-Fitting for TNOs using a Single-Epoch Parallax Measurement from L2}
\shortauthors{Giovinazzi et al.}

\begin{document}

\title{Enhancing Ground-Based Observations of Trans-Neptunian Objects using a Single-Epoch Parallax Measurement from L2}

\author[0000-0002-0078-5288]{Mark R. Giovinazzi}
\affiliation{University of Pennsylvania \\
209 S 33rd Street \\
Philadelphia, PA 19104, USA}

\author[0000-0002-6096-1749]{Cullen H. Blake}
\affiliation{University of Pennsylvania \\
209 S 33rd Street \\
Philadelphia, PA 19104, USA}

\author[0000-0003-0743-9422]{Pedro H. Bernardinelli}
\affiliation{University of Pennsylvania \\
209 S 33rd Street \\
Philadelphia, PA 19104, USA}


\begin{abstract}

Space-based observatories at the second Sun--Earth Lagrange Point (L2) offer a unique opportunity to efficiently determine the orbits of distant solar system objects by taking advantage of the parallax effect that arises from nearly simultaneous ground- and space-based observations. Given the typical orbit of an observatory about L2, the observational baseline of $\sim1.5\times10^6$~km between L2 and Earth results in an instantaneous parallax of $\sim$10$\arcsec$--100$\arcsec$, even for the most distant of detectable trans-Neptunian objects (TNOs) in our solar system. Current ground-based strategies for measuring the orbits of TNOs are very expensive and require multiple years of observations. We show that the direct constraint on the distance to a TNO, afforded by near-simultaneous ground- and space-based observations, allows us to confidently determine orbits with as few as three ground-based observations spanning 24~hr combined with a single observational epoch from an observatory orbiting L2.

\end{abstract}



\section{Introduction} \label{sec:intro}

Trans-Neptunian objects (TNOs) play a crucial role in our understanding of the origin and evolution of the solar system. Studying their distribution and chemical composition provides insight into the formation history of the Sun, planets, as well as asteroids and centaurs. The architecture of the orbital distribution for these small, rocky bodies also yields clues regarding the dynamical history of our solar system and structure in and beyond the Kuiper Belt \citep{2008ssbn.book...43G}, as well as the potential for the existence of Planet 9 \citep{Batygin_2019}.


The first TNO discovery was reported nearly 30~yr ago by \cite{Jewitt1993}. In the following decade, a number of on-sky survey searches for outer solar system objects were launched \citep[\textit{e.g.}][]{1995AJ....109.1867J, Luu_1997, brown1998a, Gladman_1998, Trujillo_2000, Trujillo2001one, Trujillo2001two, Allen_2001}. Now, the Minor Planet Center\footnote{\url{https://www.minorplanetcenter.net}} (MPC) counts $\sim$4000 outer solar system objects beyond the orbit of Neptune (semimajor axis $a>30$~au). In the coming years, new missions will drive the discovery of these outer solar system objects at an exponential rate. The Vera C. Rubin Observatory (here after the Rubin Observatory), for example, projects to detect $\sim$30,000 TNOs in 10 years of operation \citep{Ivezi__2019}. With a 9.6$^\circ$ field of view and total sky coverage of 30,000~deg$^2$, it will survey deep, wide fields quickly and often. On sheer volume alone, the Rubin Observatory will present the TNO community with an abundance of data to foster future discoveries. 

Determining a best-fit orbital solution for a TNO with current observing capabilities is time consuming and costly. Surveys typically require many observations across several years to collect enough information to run a least-squares minimization and solve for the six orbital parameters that define the Keplerian orbit of the TNO about the Sun \citep{Bernstein_2000}. For example, 2003 $\mathrm{VB}_{12}$ (Sedna), one of the most distant ($a\sim500$~au), highly elliptical ($e\sim0.84$) obsjects known to orbit the Sun, was first reported in the literature by \cite{Brown_2004}, who published an orbit with $a=480\pm40$~au based on 28 observations spaced out over more than two years. Some 15~yr later, \cite{Becker_2018} published the discovery a new object, 2015 BP$_{519}$, which has an extreme orbit ($a\sim450$~au, $e\sim0.92$) similar to that of Sedna and determined an orbital solution with 27 observations over a timespan of 1110~days. All surveys that search for TNOs \citep[see, for example,][]{Jewitt1993, 1995AJ....109.1867J, Luu_1997, brown1998a, Gladman_1998, Trujillo_2000, Trujillo2001one, Trujillo2001two, Allen_2001, Bernstein_2004, 2005AJ....129.1117E, Fraser_2008, fuentes2008subaru, Fuentes_2009, Parker_2010, Schwamb_2010, Petit_2011, Petit_2017, Alexandersen_2016, Bannister_2016, Bannister_2018, weryk2016distant, Whidden_2019, Bernardinelli_2020,y6tno}, have the same requirement that many observations of an object across multiple oppositions are needed in order to constrain its orbit. Requiring multiple oppositions' worth of data runs the risk of losing the object on the sky before properly measuring its orbit. To account for this, it is common for surveys to make multiple observations per epoch at a cadence near one hour, such that the local trajectory of the object is known.



Even larger-scale projects, which hunt for and return hundreds of classical belt (closer-in, more circular) objects, require many observations spanning multiple years. Among the most notable of such missions in recent history are the Dark Energy Survey \citep[DES,][]{2005astro.ph.10346T, Bernardinelli_2020,y6tno}, the Canada-France Ecliptic Plane Survey \citep[CFEPS,][]{Petit_2011, Petit_2017}, and the Outer Solar Systems Origin Survey \citep[OSSOS,][]{Bannister_2016, Bannister_2018}. Respectively, DES, CFEPS, and OSSOS have published the orbits of 817, 169, and 838 outer solar system objects; each mission required multiple oppositions and many observational epochs to achieve a reliable level of precision in the orbit determinations.

The second Sun--Earth Lagrange Point (hereafter L2) is one of five positions in the Sun--Earth system which act like a gravitational saddle, allowing objects placed there to remain relatively stationary during Earth's orbit. L2 is a tantalizing location for a scientific laboratory because from its perspective the three brightest objects in the sky (Sun, Earth, Moon) are clustered to a near-singular location, which allows for a telescope's sunshields to be placed such that all major light sources are blocked. This not only helps to keep an L2 facility ultra-cool, but it also provides for some large swath of sky to be observable at all times. It has hosted a number of telescopes, \textit{e.g.} the Wilkinson Microwave Anisotropy Probe \citep[WMAP,][]{Bennett_2003}, Herschel \citep{Pilbratt_2010}, and Planck \citep{planck2005-bluebook}, and is a destination spot for future facilities, \textit{e.g.} the James Webb Space Telescope \citep[JWST,][]{Gardner_2006}, Euclid \citep{laureijs2011euclid}, and the Nancy Grace Roman Space Telescope \citep{spergel2015widefield}, hereafter the Roman Space Telescope. L2 is situated $\sim$1.5 million km from Earth, and because it is considered an unstable Lagrange Point, the facilities placed there maintain a halo orbit around it that can vary from  $\sim$1.2 to $\sim$1.7 million km from Earth.

Studies of small solar system bodies are among the key science goals outlined by future L2 facilities, namely JWST, Euclid, and the Roman Space Telescope collaborations \citep{Gardner_2006, Carry_2018, holler2018solar}. \cite{parker2015physical} forecasts (on the basis of past Spitzer and HST time allocations) that $\sim$2$\times$10$^6$~s of JWST time could be allotted to TNO science. Euclid expects to make $\sim$150,000 observations of solar system objects and could discover several tens of thousands more \citep{Carry_2018}. The Roman Space Telescope, particularly its High-Latitude Survey, will be capable of detecting new TNOs, as well \citep{holler2018solar}. These three facilities will have the ability to build upon previous surveys to study TNOs \citep[\textit{e.g.} ][]{Fraser_2014, Schwamb_2019} to measure color, examine the composition and features of reflectance spectra, and more accurately determine absolute magnitudes  \citep{parker2015physical,Carry_2018,holler2018solar, fernandezvalenzuela2020compositional}. They will also assist in recovering high-fidelity orbits, and as we will show, obtaining orbital solutions in a quick, accurate, and cost-efficient way will also prove to be an important feature of the next-generation space-based observatories' scientific capabilities. Of course, there are few measurements one can make on a TNO if its location is untrackable.

Historically, so-called trigonometric parallax measurements have been made by using the angle a star appears to subtend in the sky after the Earth has moved in its orbit around the Sun to determine the distance between the star and our solar system. However, the same technique can be orchestrated by making two simultaneous measurements of the same object from Earth and some distant space-based telescope. The instantaneous distance between observatories mimics the interval previously traversed by the Earth. The New Horizons spacecraft \citep{Fountain_2008}, with a separation from Earth of $\sim$7.5 million km, has successfully demonstrated a near-instantaneous parallax measurement on nearby stars.

L2's separation from Earth is large enough to allow for near-instantaneous, or single-epoch parallax measurements of outer solar system objects. A precise parallax estimate of a TNO is tantamount to a tight constraint on its line-of-sight distance, which if known, dramatically collapses the solution space that a bound object's orbit could exist in. \cite{2011CeMDA.109..211E} successfully demonstrated this fact by recovering the orbit of a near-Earth asteroid by simulating simultaneous observations across 31 days from two other Sun--Earth Lagrange points, L4 and L5. \cite{Rhodes_2017} discussed the use of Euclid and the Rubin Observatory to directly measure distances to solar system objects using a similar technique. The apparent motion of a TNO on a near-circular orbit is approximately linear (and its distance nearly unchanged) on short timescales, and is dominated by Earth's reflex motion \citep{Bernstein_2000}. This fact instills some flexibility on the part of scheduling a pair of simultaneous observations between a space-based facility at L2 and some ground-based observatory.


We present an analysis which simulates the detection and orbit determination of a large sample of known TNOs with the inclusion of a single epoch imaging observation from an observatory at L2. A tailored ground-based TNO survey that brackets any imaging observations from L2 with a series of ground-based observations will afford a nearly instantaneous parallax measurement, which can obtain comparable, or even enhanced orbital solutions using considerably fewer observational epochs and far less total integration time. Appendices \ref{sec:formalism} and \ref{sec:coords} provide the preferred parameterization and coordinate transformations for deriving an orbital solution from the observed position of a TNO on the sky. In Section \ref{sec:param_est}, we describe the details of the statistical analysis we implement to test the effectiveness of our simulations. Section \ref{sec:feasibility} discusses the advantages that future L2 observatories will offer the community in terms of capturing TNO science and motivates our choice of the Subaru Telescope for simulating ground-based observations of TNOs. In Sections \ref{sec:results} and \ref{sec:conclusion}, we present the results of our simulations and discuss the capabilities for TNO detection using our proposed observing strategies, as well as a concluding forecast for future space-based observatories and their potential for facilitating TNO surveys.

\section{Parameter Estimation} \label{sec:param_est}

The framework of our analysis is founded upon the formalism introduced in \cite{Bernstein_2000}. We will rely on this notation to outline our proposed strategy, and so for completeness it is described fully in Appendix \ref{sec:formalism}.

To faithfully establish an orbital solution for a given set of TNO observations, we employ the Python package \texttt{emcee} \citep{emcee} to implement a Markov Chain Monte Carlo (MCMC) analysis. This Bayesian inference framework for orbit determination requires that we make initial guesses on our parameter vector $\boldsymbol{\alpha}$ (Equation \ref{eq:abg}) and impose several assumptions about the orbits of objects.

Our telescope-centric coordinate system is aligned with the object's first detection such that the total distance between observer and object is exclusively in the $\hat{\gamma}$ direction. For this reason, $\alpha_\mathrm{guess}=\beta_\mathrm{guess}=0$ is the correct initialization. The line-of-sight estimation, $\gamma_\mathrm{L2}$, comes from our single-epoch parallax measurement made using an observatory at L2 and a ground-based facility; this is described further in Section \ref{sec:feasibility}. As for the motion terms, $\dot{\alpha}$, $\dot{\beta}$, $\dot{\gamma}$ (Equation $\ref{eq:abg}$), we can initialize these assuming a circular orbit, such that $\dot{\alpha}~=~\dot{\beta}=\dot{\gamma}=\frac{2\pi}{\sqrt{3}}\gamma^{\frac{3}{2}}$. Circular orbits do not properly describe all TNOs, but even for highly elliptical orbits, these starting points provide the MCMC walkers a reasonable enough estimate to reach the correct solution.

We do not want to be overly stringent in our parameterization of the orbit, and so we enforce very few priors. The foremost assumption when fitting for any solar system object is that its orbit must be bound, though this is not always the case \citep[\textit{e.g.}][]{2017MPEC....U..181B}. For bound orbits, the eccentricity must be $e<1$ and the energy must be $E<0$. In our telescope-centric notation, the energy of an orbit is written as

\begin{equation}
E=-\mu\gamma \left( 1-2\gamma\cos\beta_0 +\gamma^2\right)^{-1/2}+ \frac{\dot{\alpha}^2+\dot{\beta}^2+\dot{\gamma}^2}{2\gamma^2}.
\label{eq:energy}
\end{equation}

Here, $\beta_0$ is the initial solar elongation angle between the Sun, Earth, and TNO. Since ground-based observations dictate the tangent-plane motion terms $\dot{\alpha}$ and $\dot{\beta}$, and the parallax measurement yields $\gamma$, Equation \ref{eq:energy} is a direct prior on the only unconstrained parameter $\dot{\gamma}$.

We employ the standard chi-squared statistic as our goodness-of-fit metric in comparing observed positions $\theta_i$ (Equation \ref{eq:gnomonic_observables}) to gnomonic models $\theta\left(\hat{\boldsymbol{\alpha}},t_i\right)$ (Equation \ref{eq:gnomonic_abg}) and inferring best-fit parameters,

\begin{equation}
    \label{eq:chi_squared_LS}
    \begin{aligned}
    \chi^2=\sum\limits_{i}^{}\left\{\frac{\left[\theta_{x,i}-\theta_x\left(\hat{\boldsymbol{\alpha}},t_i\right)\right]^2}{\sigma_i^2} + \frac{\left[\theta_{y,i}-\theta_y\left(\hat{\boldsymbol{\alpha}},t_i\right)\right]^2}{\sigma_i^2}\right\}.
    \end{aligned}
\end{equation}

The associated astrometric uncertainties $\sigma$ are specific to the instrumental precision of the observing program, and are assumed to be circular. Since a single observation from L2 affords us an excellent estimate on the distance, we include an additional contribution to our chi-squared function (a prior on distance), similar to \cite{Bernardinelli_2020},

\begin{equation}
    \label{eq:chi_squared_distance}
    \begin{aligned}
    \chi^2_\gamma=\frac{\left(1/\gamma - 1/\gamma_\mathrm{L2}\right)^2}{\sigma_z^2}.
    \end{aligned}
\end{equation}

Here, $\sigma_z$ is the expected uncertainty in our parallactic distance estimate and is defined in Section \ref{sec:feasibility}. Equations~\ref{eq:chi_squared_LS} and \ref{eq:chi_squared_distance} are combined to form our overall chi-squared, $\tilde{\chi}^2=\chi^2+\chi^2_\gamma$. We define the logarithm of the likelihood function, $\mathrm{ln}\left(\mathcal{L}\right)$, as $\mathrm{ln}\left(\mathcal{L}\right)=\frac{1}{2}\tilde{\chi}^2$ and infer a best-fit parameter vector $\boldsymbol{\alpha}$ from simulated TNO position measurements by minimizing $-\mathrm{ln}\left(\mathcal{L}\right)$. It is then trivial, using the equations prescribed in Appendices \ref{sec:formalism} and \ref{sec:coords} to convert between $\boldsymbol{\alpha}$ and the corresponding orbital solution.

The addition of our distance term (Equation \ref{eq:chi_squared_distance}) in the overall minimization is merely a simplification of the fact that, in practice, observers will incorporate their one observation from an L2 facility by mapping the measured position from L2 to the location of the ground-based observatory. This coordinate translation will have a dependence on $\gamma$, which the MCMC will marginalize over in the usual way to recover the best-fit set of input parameters to our model. While our analysis uses Equation \ref{eq:chi_squared_distance} in the minimization as a proof of concept, we reproduced fits using the generalized coordinate transform from L2 to Earth and verified that both methods produce the same resulting fits. The results presented in Section \ref{sec:results} therefore assume a series of ground-based measurements with a known distance from Earth to an object based on the inclusion of a measurement from L2.

\section{Feasibility} \label{sec:feasibility}



We highlight three future L2 facilities (JWST, Euclid, and the Roman Space Telescope) and demonstrate that each of these observatories on its own, bracketed with several observations from the ground, is capable of making a single-epoch parallax measurement of a TNO. 

 
 
For the case of JWST, an infrared space observatory projected to launch in 2021 that will extend the discoveries of the Hubble Space Telescope (HST), the Near-Infrared Camera (NIRCam), which has wavelength coverage spanning 0.6-5.0~$\mu$m, is the preferred instrument for imaging TNOs \citep{pinillaalonso2019surface}. At the shortest wavelengths, NIRCam will offer enhanced spatial resolution by a factor of two compared to HST, in addition to increased sensitivity in the range of 0.9-1.8$~\mu$m. JWST is expected to have point source positional uncertainty better than $\sigma_\mathrm{L2}=5$~mas \citep{Gardner_2006}.

Euclid's visible instrument (VIS) can image at wavelengths between 0.55 and 0.9~$\mu$m \citep{Carry_2018}. Euclid, while having a poorer spatial resolution in comparison to JWST, has a much larger 0.57~$\mathrm{deg}^2$ field of view, but aside from calibration operations will never observe within 15$^\circ$ of the ecliptic latitude nor within 30$^\circ$ of the galactic latitude. 
 
The successor to HST and JWST, the Roman Space Telescope will host two instruments capable of imaging TNOs, the Wide Field Instrument (WFI) and the Coronagraphic Instrument (CGI). The Roman Space Telescope has a 0.28~$\mathrm{deg}^2$ field of view and will cover a wavelength range of 0.6-2.0~$\mu$m \citep{holler2018solar}. Though it does not have as large a field of view as Euclid, nor a positional uncertainty as small as that of JWST, the Roman Space Telescope will be more sensitive to fainter objects.

\begin{figure}[h]
    \centering
    \includegraphics[width=\linewidth]{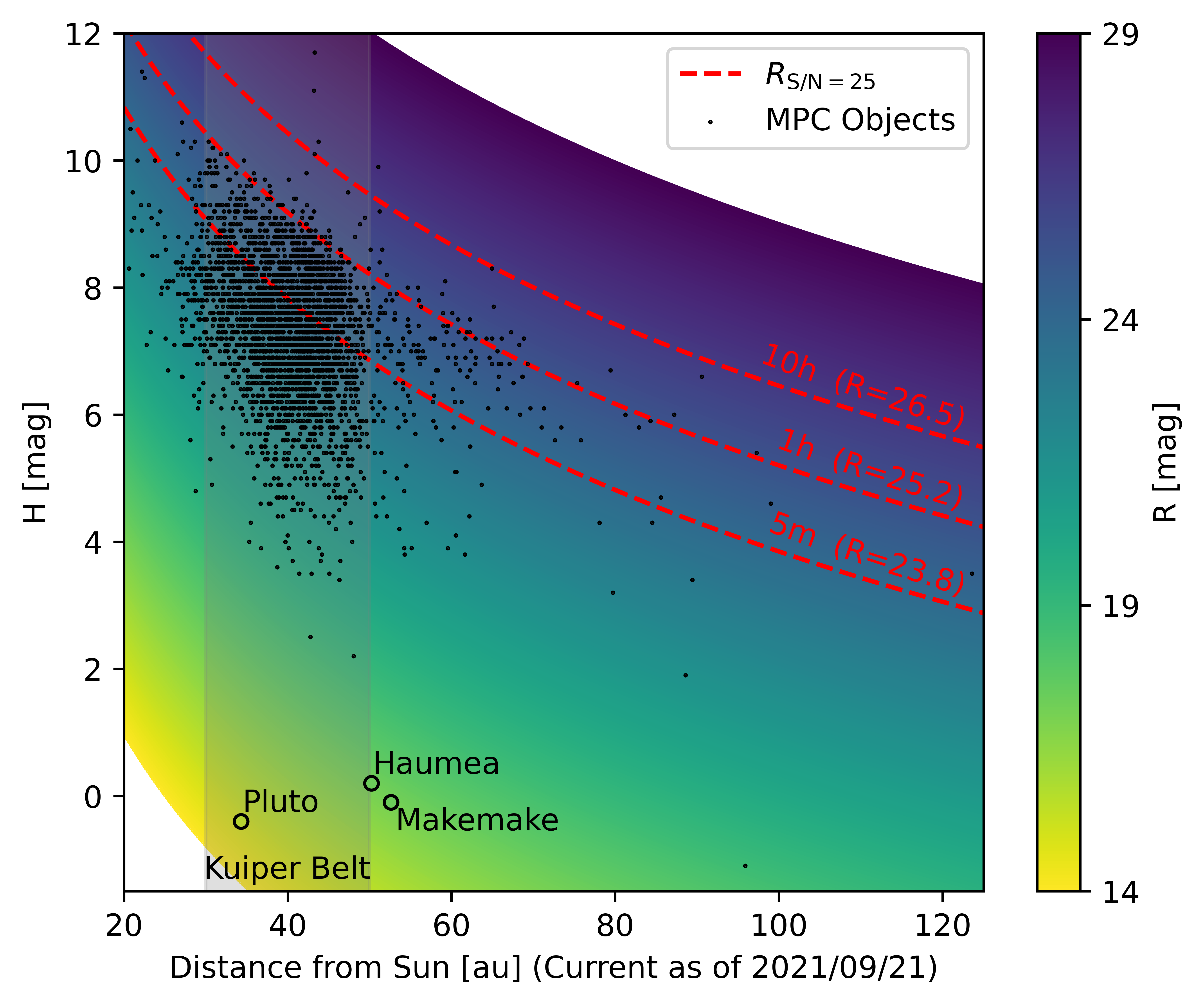} 
    \caption{Distribution of known outer solar system objects with perihelia $q>20$~au referenced in the Minor Planet Censter in terms of their absolute magnitude $H$ and current distance from the Sun. Red dashed lines give the time needed with the HSC to detect objects at corresponding $R$-band magnitudes with signal-to-noise ratio of $\mathrm{S/N=25}$ computed using the Subaru HSC Exposure Time Calculator. Pluto, Haumea, and Makemake are offered for reference.}
    \label{fig:subaru_sensitivity}
\end{figure}

For the purpose of this analysis, we assume the use of the 8.2~m Subaru Telescope at Maunakea, Hawaii. Subaru is a powerful facility for making observations of TNOs due to the large collecting area and wide-field imaging capabilities, and it has already supported several surveys for outer solar system objects \citep{2005HiA....13..773K, fuentes2008subaru, Sheppard_2016, Chen_2018, Sheppard_2019}. We simulate our observing strategies assuming the Hyper Suprime-Cam (HSC) instrument; the first two data releases from Subaru's HSC demonstrate a limiting astrometric precision of $\sigma_\mathrm{Subaru}<20$~mas \citep{Aihara_2017, Aihara_2019}; to maintain a modest estimate, we hereafter assume a measurement uncertainty from the HSC on Subaru of $\sigma_\mathrm{Subaru}=25$~mas. Preliminary results have shown that systematic searches for TNOs can be accomplished using the HSC's Subaru Strategic Program \citep{10.1093/pasj/psx082, Terai_2017, Chen_2018}. Figure \ref{fig:subaru_sensitivity} shows the distributions of absolute magnitude $H$ (defined as the visual magnitude a TNO would have if it were placed 1~au from both the Sun and Earth) versus current distance from the Sun at the time of this publication for known TNOs, and shows that most objects reported by the MPC can easily be observed within five minutes of integration on Subaru using the HSC. While recent large-scale surveys for TNOs report a steep drop in detection efficiency beyond $R\sim$23-24 \citep{Petit_2011, Bannister_2016, Bannister_2018, Bernardinelli_2020,y6tno}, a signal-to-noise ratio of $\mathrm{S/N}=25$ could be achieved from the ground on TNOs of $R\sim$24 with the HSC using only $\sim$400~s of integration according to the Subaru HSC's Exposure Time Calculator\footnote{\url{https://hscq.naoj.hawaii.edu/cgi-bin/HSC_ETC/hsc_etc.cgi}} assuming preset observing conditions. Though dozens of TNOs have already been found at fainter magnitudes than this threshold, these objects are typically discovered via more focused pencil-beam or shift-and-stack surveys \citep{Gladman_1998, Bernstein_2004, Fraser_2008, Fuentes_2009, Parker_2010}. Notably, \cite{Bernstein_2004} used this strategy to discover three TNOs at $R\gtrsim$26 using 22~ks of pointing on HST. The Rubin Observatory (first light projected in 2023) will only improve upon the sensitivity levels to which TNOs may be detected, and therefore enhance the utility of this analysis.

Observations from L2 facilities may be more expensive and difficult to schedule than those from the ground, but when we combine a series of ground-based observations with just one measurement from L2, we can instantaneously put a tight constraint on the object's distance $z$, and therefore $\gamma$, as described in Equation \ref{eq:sigma_z}; this significantly collapses the solution space and makes robust orbit determination possible with far fewer total ground-based observations and a shorter overall observational baseline. One major factor when considering the scheduling of TNO observations is the phase angle at the time of observation made between Earth, the object, and L2. Ideally, a single-epoch parallax measurement would be made when the phase angle is maximized, as a phase angle which approaches zero degrees will result in a drastically reduced parallax estimate. We simulated 100 known objects and found that TNOs will maintain 90\% of their parallactic angle between Earth and L2 within one to two months of when the phase angle is at a local maximum. Thus, there is an inherent seasonality encoded into the practicality of this analysis; objects can only have their orbits determined if they are at or near quadrature with Earth and L2.


The potential for logistical challenges in terms of scheduling nearly simultaneous measurements between ground- and space-based observatories was first detailed in the literature by \cite{2011CeMDA.109..211E}. \cite{Rhodes_2017} first emphasized the scientific synergies that will be shared by Euclid and the Rubin Observatory, which motivated \cite{snodgrass2018simultaneous} to explore at great lengths the possibility of scheduling simultaneous observations of TNOs from the two facilities. They found that, given the wide fields and rapid surveillance of both instruments, contemporaneous observations could be made supplemental to regular operation with a minimal amount of stress on either of the observatories. \cite{2019BAAS...51g..32R} also suggest the flexibility to coordinate nearly simultaneous observations between the Roman Space Telescope and Subaru via the former's GO program. In the next decade alone, there will be a multitude of opportunities to schedule single-epoch parallax measurements of TNOs between Earth and L2.

The degree of uncertainty $\sigma_z$ in the line-of-sight distance $z$ from a single-epoch parallax measurement can be estimated as a function of the total astrometric uncertainty $\sigma_\mathrm{ast}$, the distance between observatories ($d_\mathrm{L2}~=~1.5\times10^6$~km in this case), and the measured line-of-sight distance $z$. In this way,

\begin{equation}
    \label{eq:sigma_z}
    \begin{aligned}
    \sigma_z=\frac{\sigma_\mathrm{ast}}{d_\mathrm{L2}} z^2.
    \end{aligned}
\end{equation}

\begin{figure}[h]
    \centering
    \includegraphics[width=\linewidth]{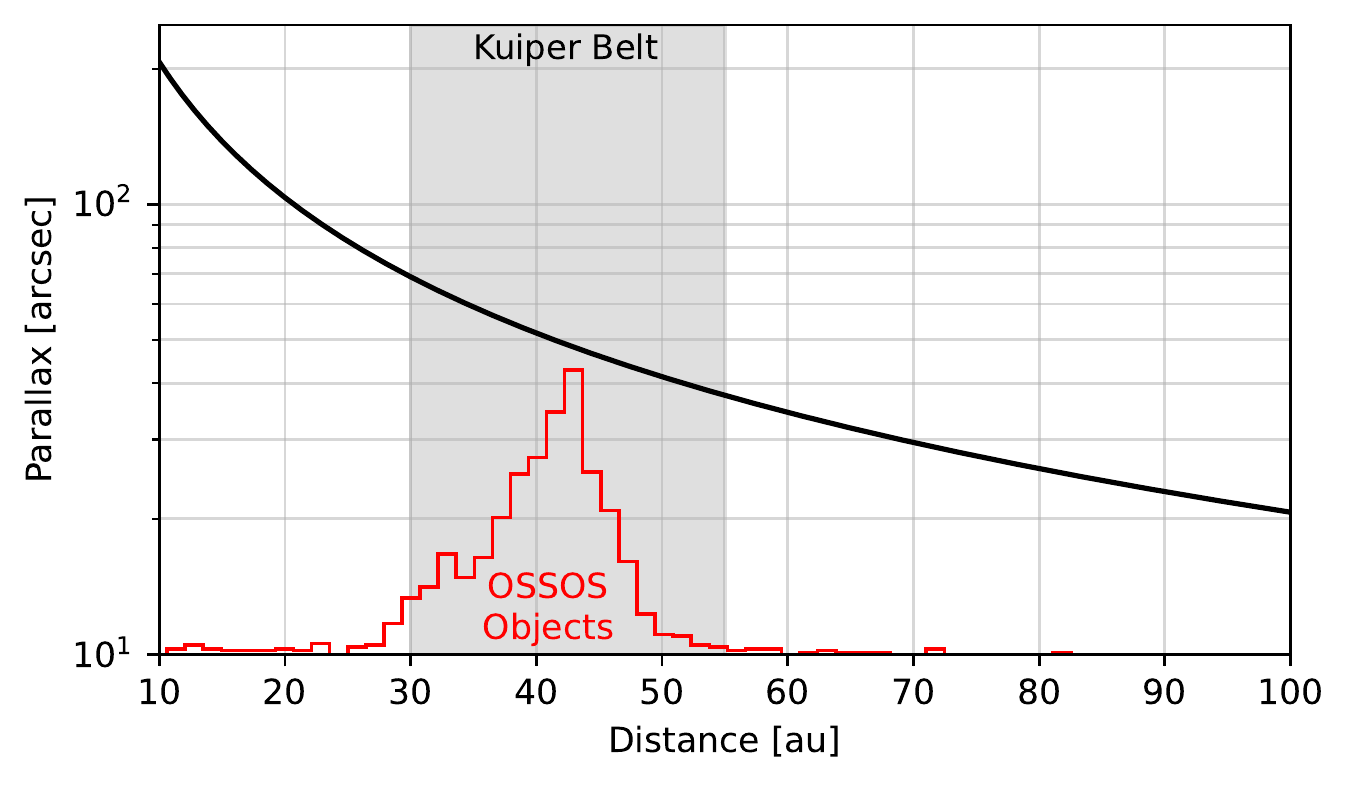} 
    \caption{Expected instantaneous parallax of a TNO given observations between L2 and a ground-based facility. Even objects in the Kuiper Belt will have parallaxes of tens of arcseconds. The distribution of discovery distances for objects published by the OSSOS campaign (which we will sample from in Section~\ref{sec:results}) is shown to demonstrate typical distances for TNOs, which are almost never observed with $z>100$~au of Earth.}
    \label{fig:parallax_plot}
\end{figure}

For well-detected objects, an astrometric precision of $\sigma_\mathrm{ast}~=~\sqrt{\sigma_\mathrm{L2}^2+\sigma_\mathrm{Subaru}^2}\sim25$~mas in the relative offset of the near-contemporaneous position from L2 and the interpolated ground-based position from three observations over 24~hr is achievable. Figure \ref{fig:parallax_plot} shows how the uncertainty on the distance $\sigma_z$ to a TNO and the measured parallax scale with the distance. Nearly all TNOs are discovered with $z<100$~au, and so achieving $\sigma_z~\sim~0.01$~au is reasonable. Recent work where there is no direct constraint on the distance assumed a prior of $\frac{\sigma_z}{z}\sim0.05$, corresponding to $\sigma_z\sim1$~au for objects at Kuiper Belt distances after multiple months of observations \citep{Bernardinelli_2020}. The value of $\sigma_z$ is an input to our Gaussian prior (Equation \ref{eq:chi_squared_distance}), as well as the standard deviation for the initialized walker distributions for $\gamma_\mathrm{L2}$ we use when initializing the \texttt{emcee} calculations. We ignore the orbit around L2 that a space-based observatory will experience and assume that its absolute position at the time of an actual observation is known.

\section{Results} \label{sec:results}

To test the effectiveness of our proposed observing strategy to measure TNO orbits including a parallax measurement using near-instantaneous observations from L2 and a ground-based facility, we simulate observations of objects published from a known TNO survey, OSSOS \citep{Bannister_2016, Bannister_2018}. OSSOS surveyed the sky for TNOs from 2013 to 2017 using the Canada-France-Hawaii Telescope. We use the HORIZONS system to generate ephemerides for these objects and produce simulated position measurements assuming observations with the HSC on the Subaru telescope.

\begin{figure}[h]
    \centering
    \includegraphics[width=\linewidth]{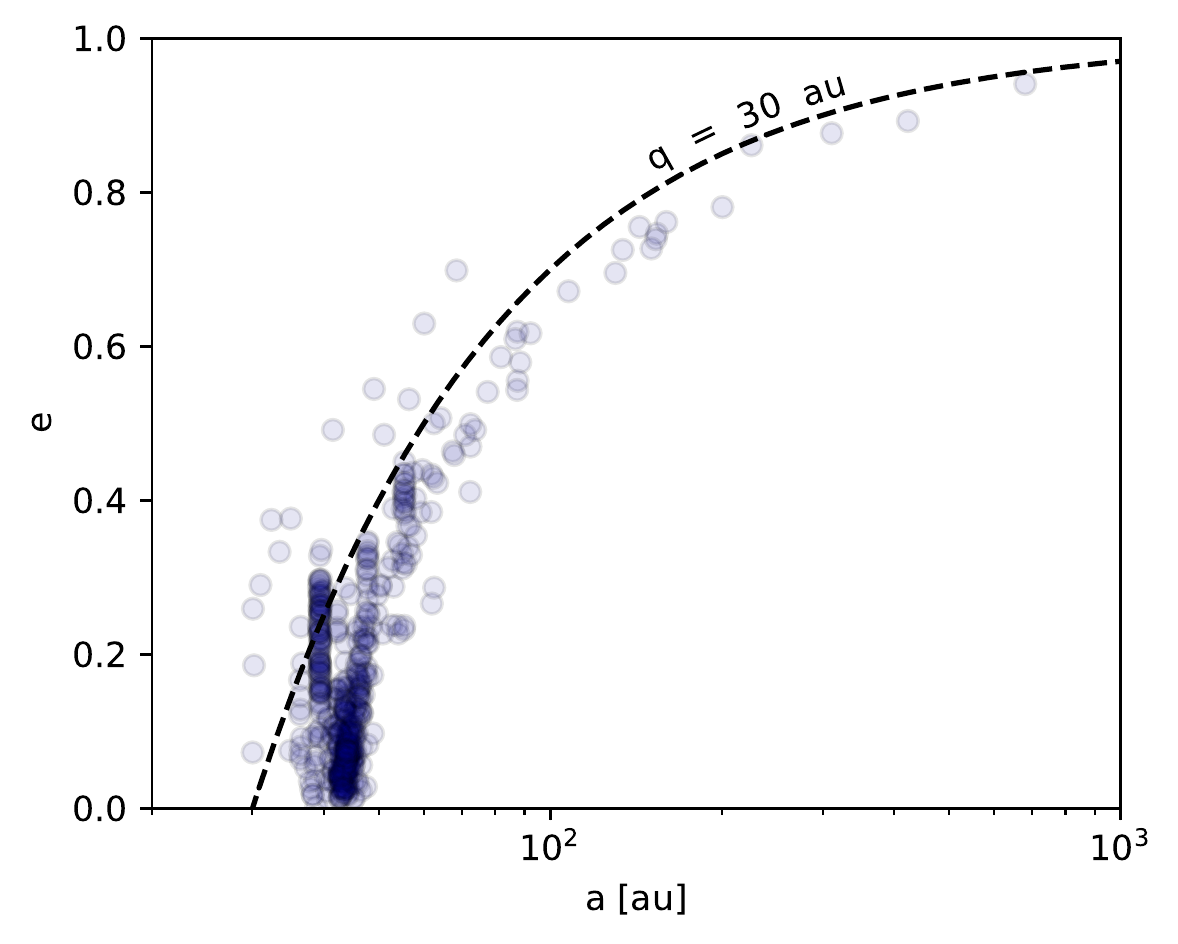} 
    \caption{Distribution of the 476 OSSOS objects that we simulate observations for.}
    \label{fig:object_sample}
\end{figure}

OSSOS objects typically required 20--60 observational epochs across 2--5 years for reliable orbit determination. We select 476 of the mission's objects (shown in Figure~\ref{fig:object_sample}) for simulated observations, each with perihelia $q~>~20$~au. To mitigate potential biases, this collection of objects cover a wide range of eccentricities and semimajor axes.

Several imaging epochs from the ground, paired with just one near-simultaneous observation from L2, is sufficient for making a parallax measurement. Given that the positions of both observatories will be very well-known, the celestial coordinates of a TNO relative to L2 can be mapped to the position on the sky as it would have been seen from the ground, and the astrometric offset due to the parallax effect can be measured relative to the same background stars. In practice, we could also modify our astrometric model of the TNO motion to explicitly include observations from two locations in the solar system, but for simplicity we have assumed that the single observation from L2 provides only a distance estimate, rather than an additional equatorial data point.



\begin{figure}[h]
    \centering
    \includegraphics[width=\linewidth]{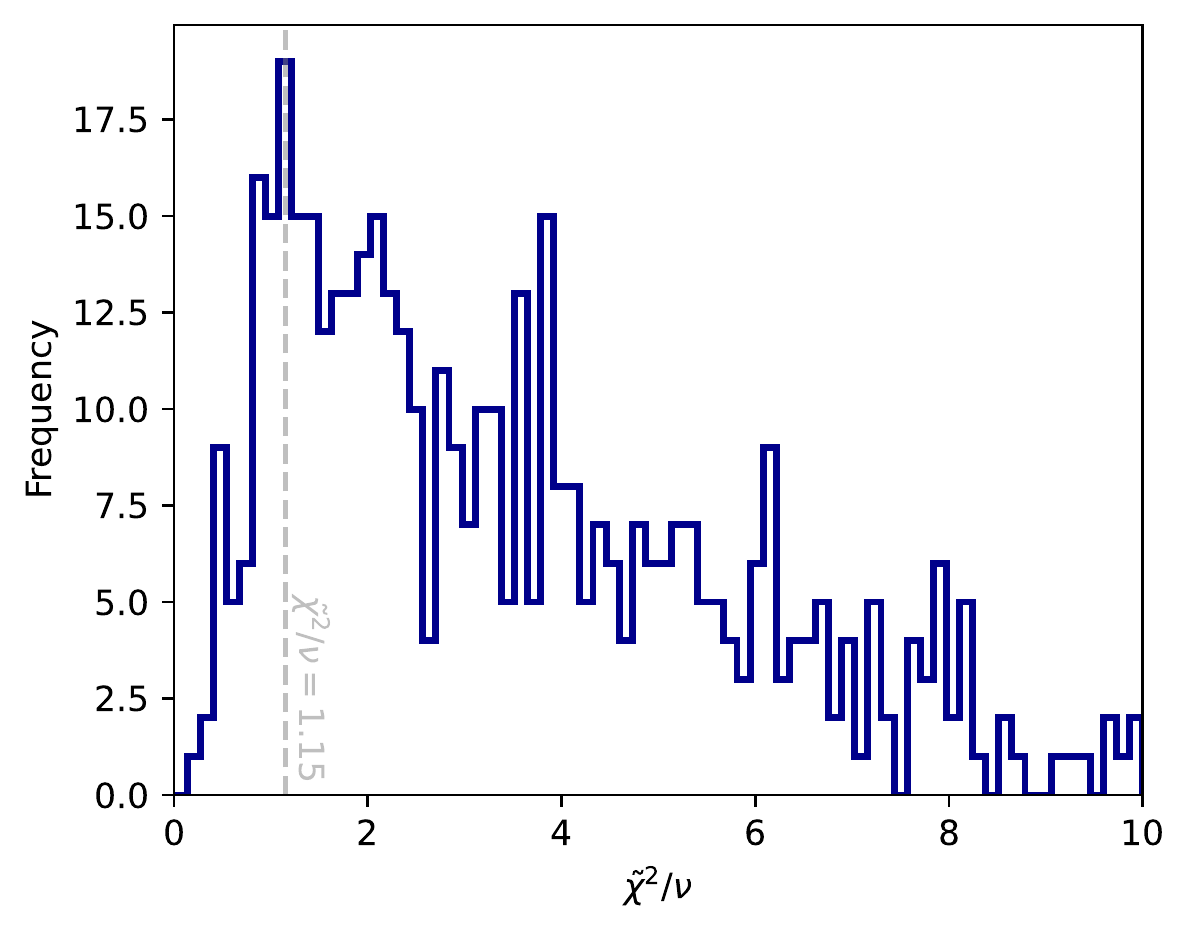} 
    \caption{Reduced $\tilde{\chi}^2$ distribution resulting from a suite of fits to a simulated population of TNOs as described in Section~\ref{sec:results}. Given three ground-based measurements and one space-based distance estimate, our fits have one remaining degree of freedom $\left(\nu=1\right)$.} 
    \label{fig:chi2_plot}
\end{figure}

We simulate ground-based measurements from Subaru of 476 known OSSOS objects over three epochs and 24~hr arcs at hours zero, six, and 24 beginning on midnight of 2015 January 1st. Since 2015 was in the middle of the OSSOS campaign, this ensures we are observing objects at or near the same locations as when they were discovered. We assume to know the position information from HORIZONS to within the prescribed uncertainty of 25 mas in both right ascension (R.A.) and declination (decl.). For each simulated set of observations, we use posteriors generated by our well-sampled, converged chains from \texttt{emcee} to infer the six best-fit telescope-centric parameters for the model as described in Section~\ref{sec:param_est}; we take the mode of these binned posterior distributions to be the best-fit values. We then fit these arcs and compare the results to the objects' true orbital parameters. In Figure \ref{fig:chi2_plot}, we show the distribution of the overall $\tilde{\chi}^2$ of the individual fits to the simulated sets of observations. All 476 simulated orbital fits successfully converged given only three ground-based measurements spanning 24 hours.


\begin{figure}[h]
    \centering
    \includegraphics[width=\linewidth]{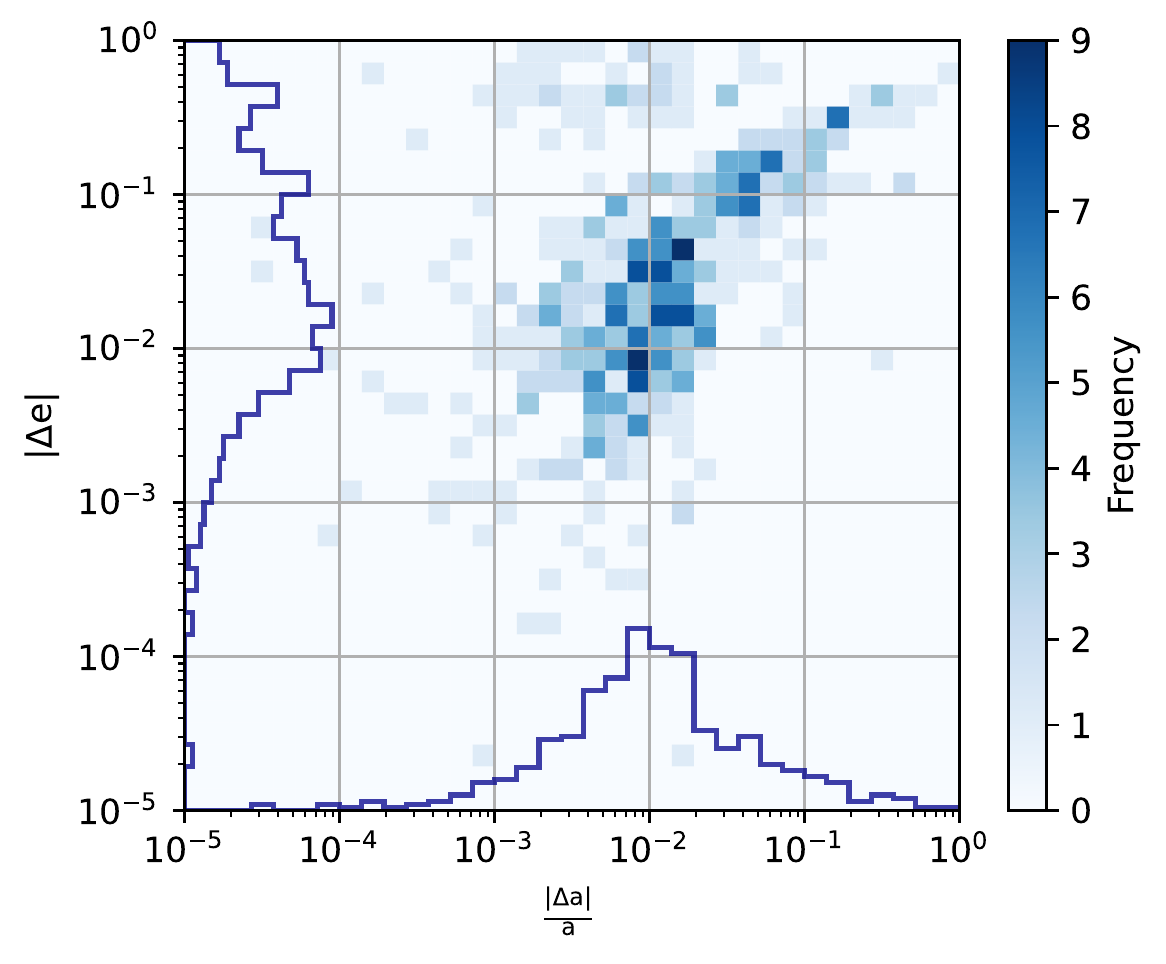} 
    \caption{Distribution of $\vert\Delta e\vert$ versus $\vert\Delta a\vert$ for our simulations. The typical TNO in our sample of objects has a semimajor axis of $a\sim$40~au.}
    \label{fig:result_plot}
\end{figure}

For a given orbital fit, the difference between our recovered and true value for semimajor axis ($\Delta a$) and eccentricity ($\Delta e$) is our chosen metric for accuracy of an orbital solution. Figure \ref{fig:result_plot} details our findings. Without an initial distance estimate, we are unable to accurately recover orbital solutions for TNOs with observations that span just 24 hours.

Our strategy correctly solves for the semimajor axis to within $\Delta a\sim1$~au and eccentricity to within $\Delta e\sim0.01$ for the majority of objects. Our result demonstrates the degeneracy which exists between eccentricity and semimajor axis in orbit recovery analyses \citep{Bernstein_2000}.

\begin{figure}[h]
    \centering
    \includegraphics[width=\linewidth]{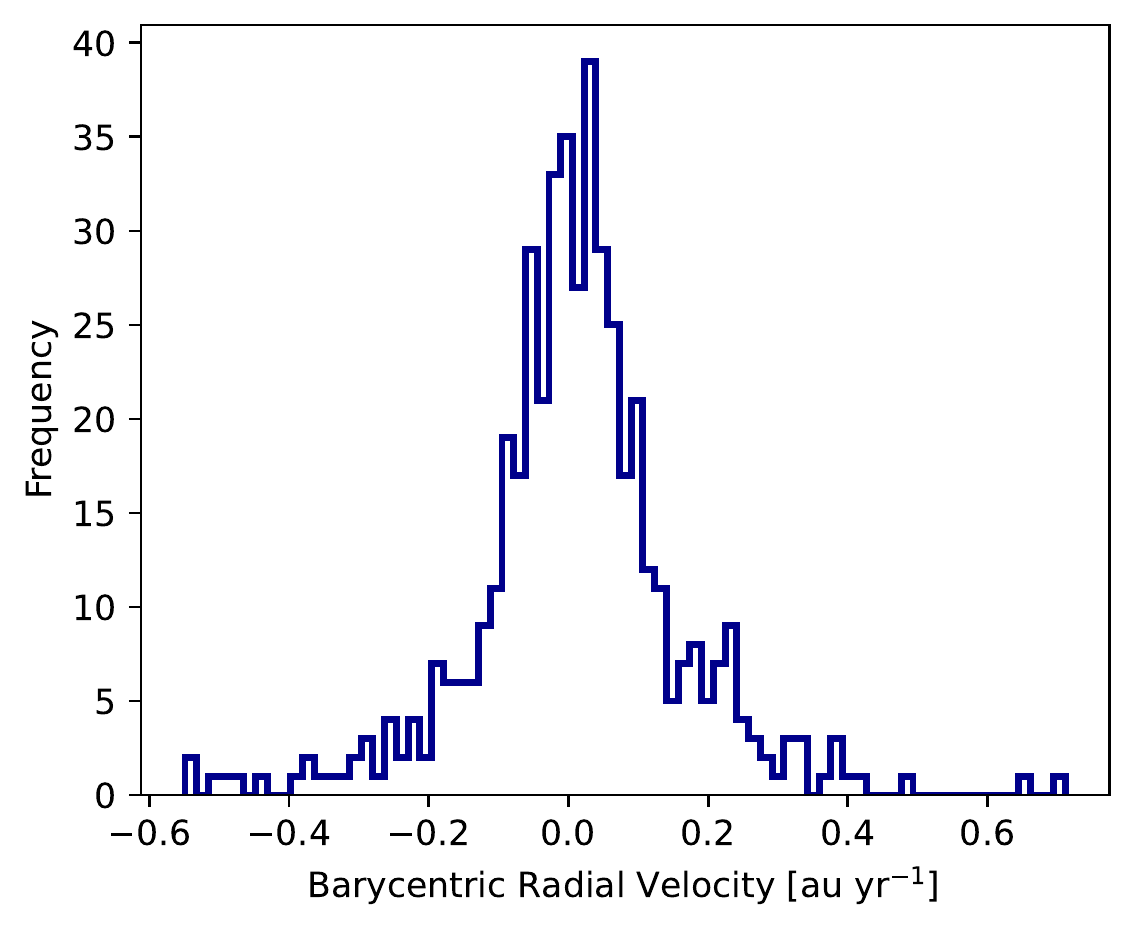} 
    \caption{Distribution of barycentric radial velocities at the time of discovery for our 476 OSSOS objects.}
    \label{fig:vrad_dist}
\end{figure}

Those objects whose $a$--$e$ pairs are poorly constrained tend to have significant motion in the line-of-sight or $\hat{\gamma}$ direction. Accurately solving for $\dot{\gamma}$ requires many months of observations, though a large fraction of TNOs have very small line-of-sight velocities and so directly constraining it is not necessary for obtaining orbital solutions for most objects. For objects with very small barycentric radial velocity, we are able to achieve orbital accuracies of $\Delta a\sim0.1$ and $\Delta e\sim0.001$. Figure~\ref{fig:vrad_dist} shows the barycentric radial velocities of the representative sample of objects we have simulated. More than 75\% of objects have absolute barycentric radial velocities $<~0.15$~au~yr$^{-1}$, for which our strategy has very little trouble obtaining the correct $a$--$e$ pair for an object's orbit. A single-epoch parallax measurement combined with just three ground-based epochs spanning 24~hr on objects with much larger absolute barycentric radial velocities, however, is less likely to obtain a correct solution. 

By obtaining accurate orbital solutions, we also attempt to recover the dynamical classification of our sample of objects. Dynamical classifications serve as a useful way of grouping together solar system objects and exploring their dynamical history \citep{2008ssbn.book...43G}. TNOs can be broadly characterized as Centaurs, classical belt objects, detached objects, scattering objects, and objects in mean motion resonances. To determine a dynamical classification for our TNOs, we use the algorithm detailed in \cite{y6tno}. For each of our objects, we generate an array of posterior orbital solutions and evolve each for 10-100~Myr alongside the Sun and outer planets using \textsc{rebound} \citep{rebound}. This evolution validates the long-term motion of the object, and will produce its maximum likelihood dynamical classification.


We find that our analysis accurately recovers the dynamical classification for $\sim$20\% of our objects. The simulation has the most success at determining classifications for objects with large semimajor axes, as there is less ambiguity between dynamical classifications, whereas the parameter space for objects with $a<50$~au requires higher orbital accuracy to confidently discern between Centaurs, classical belt objects, and scattering objects. Our recovery is is also more likely to return a correct dynamical classification for objects in mean motion resonances with Neptune, given that there is a direct relationship between orbital period and semimajor axis. Of the six Keplerian elements, $\omega$ is the most difficult to solve for, as multiple degeneracies exist within the solution space when determining the point of perihelion. Without a strong constraint on $\omega$, the orientation of the orbit is difficult to obtain; of course, this is a very minor issue for near-circular orbits. To achieve more reliable estimates of dynamical classifications, additional observations would be needed to extend the observational arcs.

\section{Conclusion} \label{sec:conclusion}


We demonstrate that by making use of a single-epoch parallax measurement, we can achieve good accuracy in determining the orbits of TNOs with just three ground-based observations over 24~hr. Our simulations indicate a typical relative uncertainty in semimajor axis of $\frac{\Delta a}{a}$ of $\sim10^{-2}$ when combining just three ground-based observations with a parallax derived from a single observation from L2. This is comparable to the level of relative uncertainty in semimajor axis achieved with much longer baselines using ground-based observations alone \citep{Bannister_2018, Bernardinelli_2020}. Reliable orbits can be determined with a shorter standard ground-based observing baseline and far less total integration time by including a single epoch from L2. Ground-based programs which closely follow the observing schedule of a space-based facility orbiting L2 will have the opportunity to jointly image hundreds of TNOs and instantaneously recover accurate orbital solutions in arc lengths of just 24~hr.

It is of interest to consider how well-timed a pair of observations between L2 and a ground-based facility would need to be in order to achieve an accurate parallax measurement. Barycentric radial distances of TNOs typically change in a measurable way on the timescale of a few weeks, and given that the single-epoch distance estimate we describe will inherit an uncertainty of $\sigma_z\sim10^{-2}$~au, it is important to obtain observations from both L2 and Earth that are close in time such that a reliable parallax estimate can be returned. Coordinating ground-based observations of an object within a week of observations from L2 should be sufficient for instantaneously and accurately measuring parallax, and thereafter constraining orbital parameters. Earth's reflex motion dominates that of the object, but Earth's trajectory is known and can be subtracted out while the object's line-of-sight motion over the course of a week is likely to be within the projected uncertainty on its distance. With three ground-based observations on an object and a near-simultaneous observation from L2, the only unconstrained parameter in the orbit is $\dot{\gamma}$. While it is possible to constrain $\dot{\gamma}$ by making two distance estimates several months apart using the single-epoch parallax method we have outlined, scheduling two simultaneous observations of thee same object from L2 across multiple months presents additional logistical challenges. For slow-moving TNOs, the object may not have moved far enough for a second, discernibly different distance to have been measured.

While a large area survey for TNOs would be prohibitively expensive, there will be programs between JWST, Euclid, and the Roman Space Telescope that will focus on outer solar system objects by targeting a number of regions of the sky. Even HST will continue hosting observing programs designed to study TNOs, therefore providing another opportunity to create an extended observational baseline capable of making parallax measurements. In fact, JWST's Cycle 1 GO Program 1568 and HST's Cycle 29 GO Program 16720 are already scheduled for nearly contemporaneous observations (2022 May-June) in the same region of sky. Optimally, grousnd-based facilities can be scheduled for nearly contemporaneous observations based on these L2 programs. In this way, a very modest investment of ground-based observing time makes it possible to determine reliable orbits for any TNO in the field of view for spaced-based observatories at L2 given that the object is bright enough to be observed from large ground-based facilities such as Subaru using the HSC.


\section{Acknowledgements}

Mark R. Giovinazzi is supported by a National Science Foundation Graduate Research Fellowship. We are grateful to Gary Bernstein for his insightful comments on best practices for fitting the orbits of TNOs, Michael McElwain and his expertise with JWST for his feedback on the utility of this analysis, and David H. Sliski for the many fruitful conversations on the details regarding this research. We also thank the anonymous referee for their insightful comments and suggestions that helped to improve this manuscript.

\appendix

\section{Formalism} \label{sec:formalism}


The telescope-centric phase space is defined by $\boldsymbol{\alpha}=\{\alpha$, $\beta$, $\gamma$, $\dot{\alpha}$, $\dot{\beta}$, $\dot{\gamma}\}$, where

\begin{equation}
    \label{eq:abg}
    \begin{aligned}
    \alpha\equiv x/z \hspace{0.5cm} \beta\equiv y/z \hspace{0.5cm} \gamma\equiv 1/z \\
    \dot{\alpha}\equiv\dot{x}/z \hspace{0.5cm} \dot{\beta}\equiv \dot{y}/z \hspace{0.5cm} \dot{\gamma}\equiv \dot{z}/z.
    \end{aligned}
\end{equation}

Here, $x$, $y$, $z$, $\dot{x}$, $\dot{y}$, and $\dot{z}$ represent the telescope-centric state vector for a given object. We emphasize that $\dot{\alpha}$, $\dot{\beta}$, and $\dot{\gamma}$ are therefore $\textit{not}$ the time derivatives of $\alpha$, $\beta$, and $\gamma$. Our telescope-centric coordinate system is centered at the initial position of the observatory, and defined such that the observer and the object are initially in the same $xz$-plane, with the total distance at $t=0$ described by the line-of-sight distance $z$. We provide transformations to move into and out of this frame in Section \ref{subsec:angular}.

The parameter vector $\boldsymbol{\alpha}$ is then used to construct a new phase space which is defined with respect to a tangent-plane, or gnomonic projection in the sky. The coordinates $\left(\theta_x, \theta_y\right)$ are defined as the Eastward direction of the J2000 ecliptic and J2000 ecliptic North, respectively, and are established in the telescope-centric reference frame as.

\begin{equation}
    \label{eq:gnomonic_abg}
    \begin{aligned}
    \theta_x\left(\boldsymbol{\alpha},t\right) = \frac{\alpha+\dot{\alpha}t+\gamma g_x\left(\boldsymbol{\alpha},t\right)-\gamma x_\mathrm{E}\left(t\right)}{1+\dot{\gamma}t+\gamma g_z\left(\boldsymbol{\alpha},t\right)-\gamma z_\mathrm{E}\left(t\right)} \\
    \theta_y\left(\boldsymbol{\alpha},t\right) = \frac{\beta+\dot{\beta}t+\gamma g_y\left(\boldsymbol{\alpha},t\right)-\gamma y_\mathrm{E}\left(t\right)}{1+\dot{\gamma}t+\gamma g_z\left(\boldsymbol{\alpha},t\right)-\gamma z_\mathrm{E}\left(t\right)}.
    \end{aligned}
\end{equation}

Above, ($x_\mathrm{E}$, $y_\mathrm{E}$, $z_\mathrm{E}$) is the position of Earth in telescope-centric coordinates relative to its initial position at $t=0$, and $\boldsymbol{g}=\left(g_x,\; g_y,\; g_z\right)$ is the gravitational perturbation term, defined as

\begin{equation}
    \label{eq:grav_perturbation}
    \begin{aligned}
    \boldsymbol{g}\left(t_0\right)=\dot{\boldsymbol{g}}\left(t_0\right)=0,\quad \ddot{\boldsymbol{g}}\left(t\right)=-\sum\limits_{i}^{}GM_i\frac{\boldsymbol{x}\left(t\right)-\boldsymbol{x}_i\left(t\right)}{\lvert\boldsymbol{x}\left(t\right)-\boldsymbol{x}_i\left(t\right)\rvert^3}.
    \end{aligned}
\end{equation}

We hereafter define $\mu\equiv\sum_{i}^{}GM_i$ to be the standard gravitational parameter summed over the Sun and all eight planets. \cite{2018AJ....156..135H} show that the linear gravitational perturbation term, written in Equation \ref{eq:linear_grav_perturbation} as $\gamma\boldsymbol{g}\left(t\right)$, can be expressed as

\begin{equation}
    \label{eq:linear_grav_perturbation}
    \begin{aligned}
    \gamma \boldsymbol{g}\left(t\right) \approx -\frac{1}{2}\mu\gamma^3 t^2 \begin{pmatrix}\alpha \\  \beta \\1   \end{pmatrix} - \frac{1}{6}\mu\gamma^3 t^3 \begin{pmatrix}\dot{\alpha} \\\dot{\beta} \\\dot{\gamma}\end{pmatrix}- 3\dot{\gamma}\begin{pmatrix}\alpha \\\beta \\ 1\end{pmatrix}.
    \end{aligned}
\end{equation}

\section{Coordinate Transformations} \label{sec:coords}

\subsection{Angular Coordinates} \label{subsec:angular}


In practice, the gnomonic projection is built from equatorial (eq) coordinates, right ascensions $\left(\mathrm{R.A.}\right)$ and declinations $\left(\mathrm{decl.}\right)$. They are translated into ecliptic latitude~$\left(l\right)$ and longitude $\left(b\right)$ as

\begin{equation}
    \label{eq:ecliptic_lb}
    \begin{aligned}
    \sin b = \cos\epsilon\sin \mathrm{\left(decl.\right)}-\sin\epsilon\cos \mathrm{\left(decl.\right)}\sin \mathrm{\left(R.A.\right)} \\
    \tan l = \frac{\cos\epsilon\cos \mathrm{\left(decl.\right)} \sin \mathrm{\left(R.A.\right)} + \sin\epsilon\sin \mathrm{\left(decl.\right)}}{\cos \mathrm{\left(decl.\right)} \cos \mathrm{\left(R.A.\right)}}.
    \end{aligned}
\end{equation}

We assume an obliquity of $\epsilon~=~23.44^\circ$, which is consistent with the JPL HORIZONS System\footnote{\url{https://ssd.jpl.nasa.gov/?horizons}}. From our ecliptic latitude and longitude, we map to our gnomonic coordinates such that

\begin{equation}
    \label{eq:gnomonic_observables}
    \begin{aligned}
    \theta_x = \frac{\cos b \sin\left(l-l_0\right)}{\sin b_0 \sin b + \cos b_0 \cos b \cos\left(l-l_0\right)} \\
    \theta_y = \frac{\cos b_0 \sin b - \sin b_0 \cos b \cos\left(l-l_0\right)}{\sin b_0 \sin b + \cos b_0 \cos b \cos\left(l-l_0\right)}.
    \end{aligned}
\end{equation}

The variables $l_0$ and $b_0$ are reference coordinates indicating the initial values of a trajectory's ecliptic latitude and longitude. These tangent-plane coordinates derived from observation are reconciled with those of Equation~\ref{eq:gnomonic_abg} to infer an orbital solution using an approach described in Section \ref{sec:param_est}. To obtain the input parameters for Equation \ref{eq:gnomonic_abg}, we require transformations between ecliptic (ec) and telescope-centric (T) coordinates. To do this, we invoke a rotation following the work of \cite{Bernstein_2000}, where

\begin{equation}
    \label{eq:ecliptic_to_telescopecentric}
    \scalebox{0.94}{$ \begin{pmatrix}x \\ y \\ z \end{pmatrix}_\mathrm{T}\begin{pmatrix} -\sin l_0  & \cos l_0 & 0 \\ -\cos l_0 \sin b_0 & -\sin l_0\sin b_0 & \cos b_0 \\ \cos l_0\cos b_0 & \sin l_0 \cos b_0  & \sin b_0 \end{pmatrix} \begin{pmatrix} x-x_0 \\ y-y_0 \\ z-z_0 \end{pmatrix}_\mathrm{ec}$.}
\end{equation}

The inverse of this transformation is then written as

\begin{equation}
    \label{eq:telescopecentric_to_ecliptic}
    \scalebox{0.91}{$  \begin{pmatrix} x-x_0 \\ y-y_0 \\ z-z_0 \end{pmatrix}_\mathrm{ec} = \begin{pmatrix} -\sin l_0 & -\cos l_0\sin b_0 & \cos l_0\cos b_0 \\  \cos l_0  & -\sin l_0\sin b_0 & \sin l_0 \cos b_0 \\ 0 & \cos b_0 & \sin b_0\end{pmatrix}   \begin{pmatrix}  x \\   y \\   z   \end{pmatrix}_\mathrm{T}$.}
\end{equation}

Therefore, to recover the ecliptic position of an object ($x_\mathrm{ec}$, $y_\mathrm{ec}$, $z_\mathrm{ec}$), we must add back in the original ecliptic position of the observatory at time $t_0$.

\subsection{Keplerian Elements} \label{subsec:keplarian}

Once an object's state vector is recovered in the ecliptic reference frame, there are a series of conversions we can apply to translate between Cartesian coordinates and the Keplerian orbital elements. To construct the full orbital basis set given an instantaneous position $\boldsymbol{r}$ and velocity $\boldsymbol{v}$ of an orbiting body, we define the angular momentum vector $\boldsymbol{h}=\boldsymbol{r}\times\boldsymbol{v}$, as well as a vector which points in the direction of the ascending node $\boldsymbol{n}= \hat{z}\times\boldsymbol{h}$. Using those, we follow a recipe for computing semimajor axis $a$, eccentricity $e$, inclination $i$, argument of perihelion $\omega$, and longitude of the ascending node $\Omega$, respectively, with 

\begin{equation}
    \label{eq:a}
    \begin{aligned}
    a = \left( \frac{2}{\lvert\boldsymbol{r}\rvert} - \frac{v^2}{\mu} \right)^{-1}
    \end{aligned}
\end{equation}

\begin{equation}
    \label{eq:e}
    \begin{aligned}
    \boldsymbol{e}=\frac{\boldsymbol{v}\times\boldsymbol{h}}{\mu} - \frac{\boldsymbol{r}}{\lvert \boldsymbol{r} \rvert},\quad e=\lvert\boldsymbol{e}\rvert
    \end{aligned}
\end{equation}

\begin{equation}
    \label{eq:i}
    \begin{aligned}
    i= \cos^{-1}\left(\frac{h_z}{\lvert \boldsymbol{h} \rvert}\right)
    \end{aligned}
\end{equation}

\begin{equation}
    \label{eq:omega}
    \begin{aligned}
    \omega = \cos^{-1}\left(\frac{\boldsymbol{n}\cdot\boldsymbol{e}}{\lvert \boldsymbol{n} \rvert \lvert \boldsymbol{e} \rvert}\right)
    \end{aligned}
\end{equation}  

\begin{equation}
    \label{eq:Omega}
    \begin{aligned}
    \Omega = \cos^{-1}\left(\frac{n_x}{\lvert\boldsymbol{n}\rvert}\right).
    \end{aligned}
\end{equation}

\bibliography{sample63}{}
\bibliographystyle{aasjournal}

\end{document}